\documentclass[twocolumn,english]{revtex4}
\usepackage{mathptmx}

\usepackage[T1]{fontenc}
\usepackage[latin9]{inputenc}
\usepackage{graphicx}
\usepackage{amssymb}
\usepackage{esint}

\makeatletter
\@ifundefined{textcolor}{}
{%
 \definecolor{BLACK}{gray}{0}
 \definecolor{WHITE}{gray}{1}
 \definecolor{RED}{rgb}{1,0,0}
 \definecolor{GREEN}{rgb}{0,1,0}
 \definecolor{BLUE}{rgb}{0,0,1}
 \definecolor{CYAN}{cmyk}{1,0,0,0}
 \definecolor{MAGENTA}{cmyk}{0,1,0,0}
 \definecolor{YELLOW}{cmyk}{0,0,1,0}
 }

\def\aap{{Astronomy and Astrophys.}}	

\def\araa{{ARA\&A}}

\def\mnras{{MNRAS}}

\def\apj{{\it Astrophys. J.\ }}
		
\def\apjl{{\apj\ \it Lett.\ }}

\def\physrep{{\it Phys.~Rep.\ }}
\def\mnras{{\it Mon. Not. R. Astron. Soc.\ }}

\def\nat{{\it Nature}}

\usepackage{hyperref}

\makeatother

\usepackage{babel}

\begin{document}

\title{Mechanism for spectral break in cosmic ray proton spectrum from Supernova remnant W44}

\author{M.A. Malkov$^{1}$, P.H. Diamond$^{1}$ and R.Z. Sagdeev$^{2}$}

\affiliation{$^{1}$CASS and Department of Physics, University of California,
San Diego, La Jolla, CA 92093}

\affiliation{$^{2}$University of Maryland, College Park, Maryland 20742-3280,
USA}
\begin{abstract}
Recent observations of the supernova remnant W44 by the \emph{Fermi
}spacecraft observatory strongly support the idea that the bulk of
galactic cosmic rays is accelerated in such remnants by a Fermi mechanism,
also known as diffusive shock acceleration. However, the W44 expands
into weakly ionized dense gas, and so a significant revision of the
mechanism is required. In this paper we provide the necessary modifications
and demonstrate that strong ion-neutral collisions in the remnant
surrounding lead to the steepening of the energy spectrum of accelerated
particles by \emph{exactly one power}. The spectral break is caused
by Alfven wave evanescence leading to the fractional particle losses.
The gamma-ray spectrum generated in collisions of the accelerated
protons with the ambient gas is also calculated and successfully fitted
to the Fermi Observatory data. The parent proton spectrum is best
represented by a classical test particle power law $\propto E^{-2}$,
steepening to $E^{-3}$ at $E_{br}\approx7GeV$ due to deteriorated
particle confinement.
\end{abstract}
\maketitle
Ongoing \emph{direct }observations of a number of galactic supernova
remnants (SNRs) in the X- and gamma-ray bands supported by the radio,
optical and UV data \citep{Reynolds08ARA,Raymond01} seem to rapidly
close in on the century long problem of the cosmic ray (CR) origin.
Overall, the observations favor the diffusive shock acceleration (DSA,
operating in SNR shocks) as a mechanism for the production of galactic
CRs. However, there are at least two critical questions that observations
pose to the theory. First, what exactly is the form of the spectrum
that the theory predicts? Second, when do we really see the proton
(i.e., the primary CR component) emission and when do we see just
a contaminating but radiatively more efficient electron component?

Several decades of studies of the background galactic CRs, preceding
direct observations (that became available only recently) demonstrated
that already a simple test particle version of the DSA (leading to
a power-law $\propto E^{-2}$ particle energy distribution) reproduces
the form of the energy spectrum reasonably well. It is possible, however,
that a somewhat steeper power-law would better accommodate the $E^{-2.7}$
measured spectrum \citep{Gaisser98,Hillas05} (see also the Discussion
section below), even though this difference can be attributed to CR
propagation losses (also not completely certain). Nevertheless, the
\emph{direct observations }of emission coming from particles accelerated
in SNR shocks often indicate significantly steeper than $E^{-2}$
spectra \citep{Ahar06RXJ}. Also the recent \emph{Fermi} Large Area
Telescope (LAT) observations of the SNR W44 suggest that the spectrum
of the gamma ray producing protons is substantially steeper at high
energies than the DSA predicts \citep{Abdo10W44full}. It should be
noted that a similar discrepancy has already been found in the high
energy gamma ray spectrum of the SNR RXJ 1713 measured by the CANGAROO
\citep{Enomoto02} and HESS \citep{Ahar06RXJ} atmospheric Cerenkov
telescopes. The simple DSA prediction was even used as an argument
against the hadronic origin of the gamma emission on the ground that
if the high energy part of the spectrum was extrapolated (with the
DSA index 2) to lower energies, the emission there would be unacceptably
high, e.g. \citep{Pohl02}. Motivated by this inconsistency with the
DSA theory in particular, we have previously published a suggestion
of the mechanism for the spectral break \citep{MDS05}, which can
resolve the discrepancies. We believe that a similar phenomenon is
also applicable to the recent observations of the SNR W44.

In particular, we show below that ion-neutral collisions in a molecular
cloud adjacent to the remnant steepen the energy spectrum of accelerated
particles. We calculate the gamma-ray spectrum resulting from the
collisions of accelerated protons with the molecular cloud material
and fit the result to the Fermi Observatory data.

\section*{Results}

\paragraph*{Mechanism for the spectral break}

The physics of the spectral break is very simple. When a SNR shock
approaches a molecular cloud (MC) or a pre-supernova swept-up shell,
confinement of accelerated particles generally deteriorates. Due to
the particle interaction with magnetic turbulence, confinement requires
scales similar to the particle gyroradius \citep{Drury83,BlandEich87}.
While the waves are in a strongly ionized (closer to the shock) medium
they propagate freely in a broad frequency range at the Alfven speed
$V_{A}=B/\sqrt{4\pi\rho_{i}}$ with the frequencies $\omega=kV_{A}$.
Here $k$ is the wave number (assumed parallel to the local field
$\mathbf{B}$) and $\rho_{i}$ is the ion mass density. As long as
the Alfven wave frequency is higher than the ion-neutral collision
frequency $\nu_{in}$, the waves are weakly damped. When, on the other
hand, the ion-neutral collision frequency is higher (deeper into the
cloud), neutrals are entrained by the oscillating plasma and the Alfven
waves are also able to propagate, albeit with a factor $\sqrt{\rho_{i}/\rho_{0}}<1$
lower speed, where $\rho_{0}$ is the neutral density. The propagation
speed reduction occurs because every ion is now {}``loaded'' with
$\rho_{0}/\rho_{i}$ neutrals. In between these two regimes Alfven
waves are heavily damped and even disappear altogether for sufficiently
small $\rho_{i}/\rho_{0}\ll0.1$. The evanescence wave number range
is then bounded by $k_{1}=\nu_{in}/2V_{A}$ and $k_{2}=2\sqrt{\rho_{i}/\rho_{0}}\nu_{in}/V_{A}$.
These phenomena have been studied in detail in \citep{KulsrNeutr69,ZweibelShull82},
and specifically in the context of the DSA in \citep{VoelkNeutrDamp81,DruryNeutral96,RevilleNeutr08}.
Now we turn to their impact on the particle confinement and emissivity.

\begin{figure}
\includegraphics[scale=1.2]{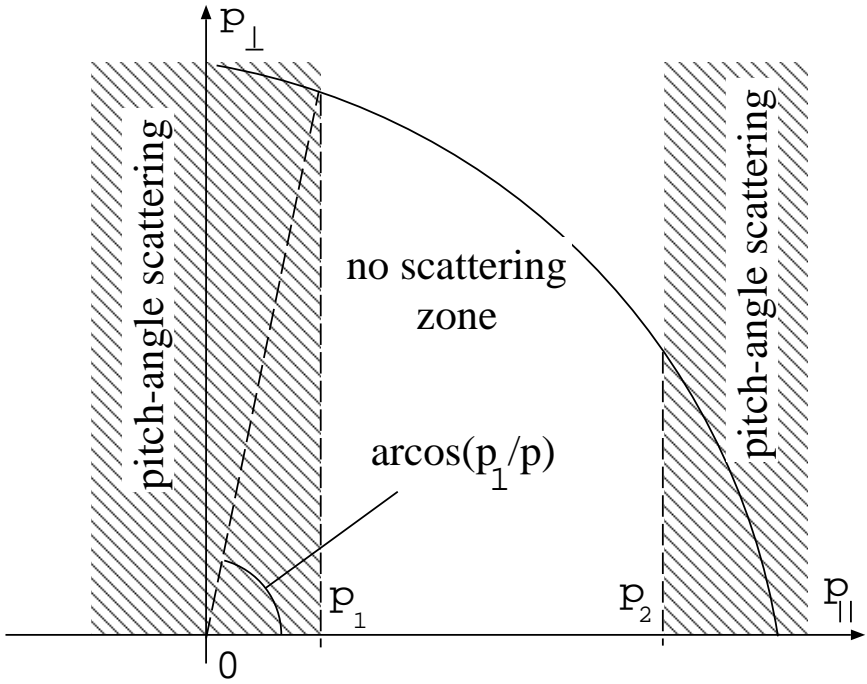}
\caption{\textbf{Momentum space of accelerated protons. }Particle scattering
zones on the $\left(p_{\parallel},p_{\perp}\right)$- plane of momentum
space. Pr in the stripes $p_{1}<\left|p_{\parallel}\right|<p_{2}$
are not scattered by waves (see text). Therefore, particles from the
domains $\left|p_{\parallel}\right|>p_{2}$ maintain their propagation
direction and promptly escape from the dense gas region. \label{fig:Scattering-zone-of}}

\end{figure}

In the frame work of a quasilinear wave-particle interaction the wave
number $k$ is approximately related to the parallel (to the magnetic
field) component of the particle momentum $p_{\parallel}$ by the
cyclotron resonance condition $kp_{\parallel}/m=\pm\omega_{c}$ where
the (non-relativistic) gyro-frequency $\omega_{c}=eB/mc$. Therefore,
the frequency range where the waves cannot propagate may be conveniently
translated into the parallel momentum range

\begin{equation}
p_{1}<\left|p_{\parallel}\right|<p_{2}\label{eq:ineq}\end{equation}
with 

\begin{equation}
p_{1}=2V_{A}m\omega_{c}/\nu_{in},\;\; p_{2}=\frac{p_{1}}{4}\sqrt{\rho_{0}/\rho_{i}}>p_{1},\label{eq:p12}\end{equation}
That a spectral break must form at the photon energy corresponding
to the particle momentum $p=p_{1}=p_{br}$, can be readily understood
from Fig.\ref{fig:Scattering-zone-of}. The 'dead zones' $p_{1}<\left|p_{\parallel}\right|<p_{2}$
imply that particles with $\left|p_{\parallel}\right|>p_{1}$ do not
turn around (while moving along the magnetic field) and escape from
the region of CR-dense gas collisions at a $p_{\parallel}/p$ fraction
of the speed of light. This happens in the region of enhanced gamma
radiation. Therefore, an initially isotropic distribution of accelerated
particles fills only a slab in momentum space $\left|p_{\parallel}\right|<p_{1}$
and becomes highly anisotropic (a 'pancake' distribution). What matters
for the integral emission, however, is a locally isotropic component
$\overline{f}$ of this new proton distribution. It can be introduced
by re-averaging the 'pancake' ($\left|p_{\parallel}\right|<p_{1}$)
distribution in pitch angle, $\overline{f}\left(p\right)\equiv\intop_{0}^{1}f\left(p,\mu\right)d\mu$,
and is readily obtained assuming that particles remaining in the dense
gas (those with $\left|p_{\parallel}\right|<p_{1}$) maintain their
isotropic pitch-angle distribution, i.e. 

\begin{eqnarray}
\overline{f}\left(p\right) & = & \intop_{0}^{\mu_{1}}f_{0}\left(p\right)d\mu=\left\{ \begin{array}{cc}
\left(p_{1}/p\right)f_{0}\left(p\right), & p\ge p_{1}\\
f_{0}\left(p\right), & p<p_{1}\end{array}\right.\label{eq:fbar}\end{eqnarray}
where $f_{0}\left(p\right)$ is the initial (isotropic) distribution
function and $\mu$ is the cosine of the pitch angle (see Fig.\ref{fig:Scattering-zone-of}),
$\mu_{1}=\min\left\{ p_{1}/p,1\right\} $. Thus, the slope of the
particle momentum distribution becomes steeper by exactly one power
above $p=p_{1}\equiv p_{br}$. In particular, any power-law distribution
$\propto p^{-q}$, upon entering an MC, turns into $p^{-q-1}$ at
$p\ge p_{br}$, and preserves its form at $p<p_{br}$.

Now the question is whether particle escape from the MC can quench
the acceleration process itself. In principle it can, since also the
particles from the interval $\left|p_{\parallel}\right|<p_{1}$ cannot
return to the shock and continue acceleration. Instead, by slowly
leaking to the part of the momentum space with $\left|p_{\parallel}\right|>p_{1}$,
they should ultimately escape as well. This would certainly be the
case if the MC were filling the entire shock precursor. However, MCs
are known to be clumpy \citep{CrutcherMC99,Chev03,Pariz04}, and may
fill only a small fraction of the precursor. In this case the acceleration
process continues almost unimpeded, the accelerated protons illuminate
the 'cloudlets' and make them visible in $\gamma$-rays due to the
high density target material.

\paragraph*{Break momentum}

While the one power spectral break in the pitch-angle averaged particle
distribution seems to be a robust environmental signature of a weakly
ionized medium into which the accelerated particles propagate, the
break momentum remains uncertain. According to eq.(\ref{eq:p12}),
$p_{br}$ ($\equiv p_{1}$) depends on the magnetic field strength
and ion density as well as on the frequency of ion-neutral collisions,
$\nu_{in}=n_{0}\left\langle \sigma V\right\rangle $. Here $\left\langle \sigma V\right\rangle $
is the product of the collision cross-section and collision velocity
averaged over the thermal distribution. Using an approximation of
\citep{DraineMcKee93,DruryNeutral96} for $\left\langle \sigma V\right\rangle $,
$p_{br}$ can be estimated as \begin{equation}
p_{br}/mc\simeq10B_{\mu}^{2}T_{4}^{-0.4}n_{0}^{-1}n_{i}^{-1/2}.\label{eq:p1}\end{equation}
Here the gas temperature $T_{4}$ is measured in the units of $10^{4}K$,
magnetic field $B_{\mu}$ -in microgauss, $n_{0}$ and $n_{i}$ (number
densities corresponding to the neutral/ion mass densities $\rho_{0}$
and $\rho_{i}$) \--in $cm^{-3}$. Note that the numerical coefficient
in the last expression may vary depending on the average ion and neutral
masses and can be higher by a factor of a few for typical molecular
cloud conditions \citep{KulsrNeutr69,NakanoMC84} than the estimate
in eq.(\ref{eq:p1}) suggests. The remaining quantities in the last
formula are also known too poorly to make an accurate independent
prediction of the position of the break in the gamma ray emission
region. Those are the regions near the blast wave where complicated
physical processes unfold. They include particle acceleration, strong
MHD turbulence (driven by particles and their interaction with ambient
gas inhomogeneities), gas ionization by shock generated UV photons,
turbulent plasma heating and even evaporation of magnetic cloudlets
\citep{ShullMcKeeMC79,DraineMcKee93,BykovMC00}. Also important may
be the ionization by the low energy CRs accelerated at the blast wave.
However, as their diffusion length is shorter than that of the particles
with $p\gtrsim p_{br}$, we may assume that they do not reach the
MC. Pre-ionization by the UV photons can also be ignored for the column
density $N>10^{19}cm^{-2}$ ahead of the shock beyond which they are
absorbed \citep{Uchiyama10}. The authors or the Ref. \citep{Uchiyama10},
using the earlier data from \citep{Reach05} have also analyzed the
parameters involved in eq.(\ref{eq:p1}) and found the above estimate
of $p_{br}$ to be in a good agreement with the spectral break position
measured by the \emph{Fermi} LAT. Nevertheless, we may run the argument
in reverse and use the \emph{Fermi} observations \citep{Abdo10W44full}
of the gamma-ray spectrum of SNR W44 to determine the break momentum
in the parent particle spectrum and constrain the parameters in eq.(\ref{eq:p1}).
Once we also know the amount of the slope variation $\Delta q$, we
can calculate the full spectrum up to the cut-off energy.

\paragraph*{Particle spectra}

To calculate the particle spectra, we need to determine the degree
of nonlinear modification of the shock structure. In principle, it
can be calculated consistently, given the shock parameters and the
particle maximum momentum, $p_{max}$. In the case of a broken spectrum,
$p_{br}$ plays the role of $p_{max}$, since this is the momentum
where the dominant contribution to the pressure of accelerated particles
comes from, thus setting the scale of the modified shock precursor.
Note that in the conventional nonlinear (NL) acceleration theory,
the cut-off momentum $p_{max}$ plays this role, because the nonlinear
spectra are sufficiently flat so as to make the pressure diverge with
momentum.

The break in the photon spectrum is observed at about $2$ GeV, which
places the break on the proton distribution at about $p_{br}\simeq7GeV/c$
\citep{Abdo10W44full}. For the strength of the break $\Delta q=1$,
the spectrum above it is clearly pressure converging, and we perform
the calculation of the shock structure and the spectrum using this
break momentum as the point of the maximum in the CR partial pressure.
Once this point is set, we can use an analytic approach for a stationary
nonlinear acceleration problem using $p_{br}$ as an input parameter. 

Apart from $p_{br}$, the nonlinear solution depends on a number of
other parameters, such as the injection rate of thermal particles
into acceleration, Mach number, the precursor heating rate and the
shock velocity $V_{s}$. Of these parameters only the latter is known
accurately, $V_{s}\approx300km/s$, the other parameters are still
difficult to ascertain. Fortunately, in sufficiently strong shocks
the solution generally tends to either stay close to the test particle
(TP) solution (leaving the shock structure only weakly modified) or
else it transitions to a strongly modified NL-solution regime. The
TP regime typically establishes in cases of moderate Mach numbers,
low injection rates and low $p_{max}$ (now $p_{br}$), while the
NL regime is unavoidable in the opposite part of the parameter space.

In the TP regime the spectrum is close to a power-law with the spectral
index 2 throughout the supra-thermal energy range. In the NL regime,
however, the spectrum develops a concave form, starting from a softer
spectrum at the injection energy, with the index $q\simeq(r_{s}+2)/(r_{s}-1)>2$,
where $r_{s}<4$ is the sub-shock compression ratio. Then it hardens,
primarily in the region $p\sim mc$, where both the partial pressure
and diffusivity of protons change their momentum scaling. The slope
reaches its minimum at the cut-off (break) energy, which, depending
on the degree of nonlinearity, can be as low as 1.5 or even somewhat
lower if the cut-off is abrupt. The question now is into which of
these two categories the W44 spectrum falls? Generally, in cases of
low maximum (or, equivalently, low spectral break $p_{br}\lesssim10$)
momentum, the shock modification is weak, so the spectrum is more
likely to be in the only slightly nonlinear, almost TP regime. On
the other hand, there is a putative indication from the electron radio
emission that their spectrum may be close to $q_{e}\approx1.75$,
which could be the signature of a moderately nonlinear acceleration
process. It should be remembered, however, that this is a global index
across the W44 remnant. There are resolved bright filaments where
a canonical $\alpha=-0.5$ spectrum, corresponding precisely to the
TP parent electron spectrum with $q_{e}=2$ is observed \citep{W44Radio07}.
Moreover, there are regions with the positive indices $\alpha\lesssim0.4$
which cannot be indicative of a DSA process without corrections for
subsequent spectral transformations such as an absorption by thermal
electrons \citep{W44Radio07}. These regions can very well contribute
to the overall spectral hardening discussed above, mimicking the acceleration
nonlinearity. Finally, secondary electrons give rise to the flattening
of the radio spectrum as well \citep{Uchiyama10}.

\begin{figure}
\includegraphics[scale=0.7]{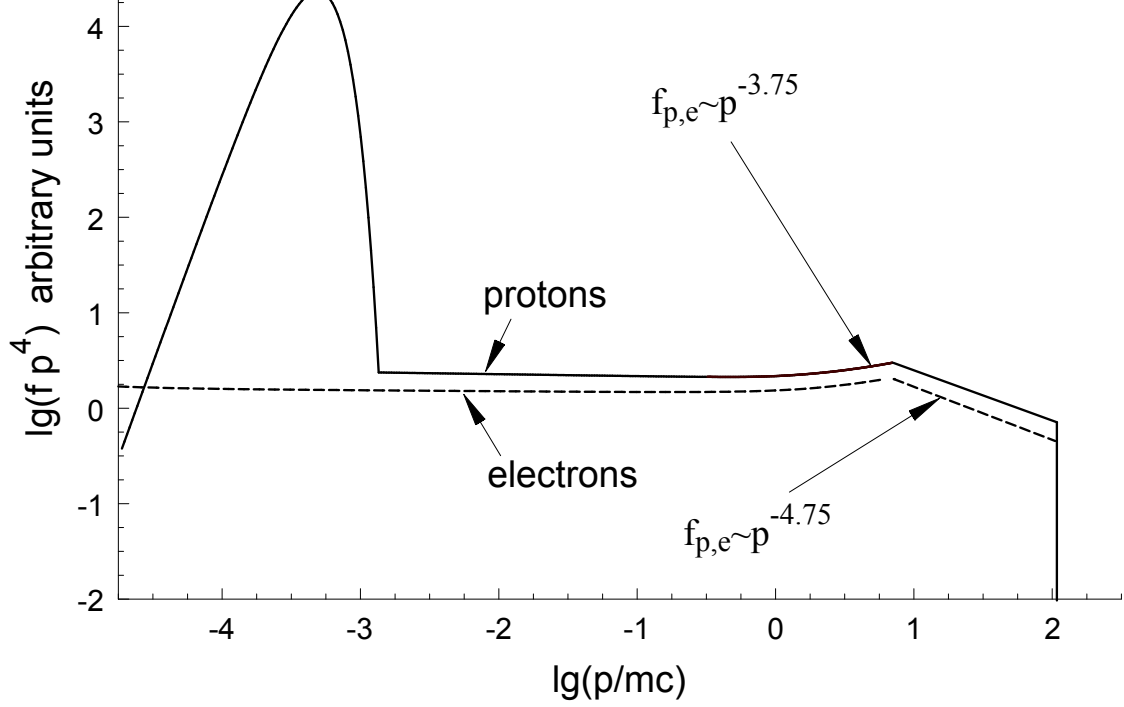}
\caption{\textbf{Spectra of accelerated protons and electrons. }The both particle
distributions are calculated for a weakly modified shock and are shown
in momentum normalization ($f\left(p\right)$ is steeper by two powers
than the spectra in energy normalization, used in the text). Both
spectra are multiplied by $p^{4}$, so that the test particle distribution
is flat. Shock parameters: acoustic Mach number $M=30$, shock velocity
$V_{s}/c=10^{-3}$, the break momentum $p_{br}\simeq7mc$. Shock pre-compression
(flow compression across the CR precursor) R=1.8, injection parameter
$\nu\simeq0.1$ {[}defined as $\nu=\left(4\pi/3\right)\left(mc^{2}/\rho V_{s}^{2}\right)\left(p_{inj}/mc\right)^{4}f\left(p_{inj}\right)$,
with $\rho$ and $V_{s}$ being the ambient gas density and the shock
speed, respectively{]}; injection momentum $p_{inj}/mc\simeq1.4\cdot10^{-3}$.\label{fig:Spectra-of-protons}}

\end{figure}

If the accelerated protons and electrons respond to the turbulence
similarly, which is almost certainly the case in the ultra-relativistic
regime, their spectra should have similar slopes there (as long as
the synchrotron losses are ignorable). In using the electron radio
spectrum as a probe for the level of acceleration nonlinearity, the
following two relations are useful. First, there is a relation between
the electron energy and the radio frequency $\nu_{MHz}=4.6\cdot B_{\mu}E_{GeV}^{2}$.
The second relation, $q_{e}=1-2\alpha$, links the spectral index
of radio emission $\alpha$ (assuming the radio flux $\propto\nu^{\alpha}$)
and the spectral index of the parent electrons $q_{e}$ (assuming
their energy spectrum $\propto E^{-q_{e}}$). Once the global radio
spectral index of W44, $\alpha\simeq-0.37$ \citep{W44Radio07} is
generated by freshly accelerated electrons in the frequency range
$74<\nu<10700$ MHz, the electrons should maintain their modified
spectrum over the energy range spanning more than one order of magnitude.
For example, assuming $B_{\mu}\simeq70$ \citep{Abdo10W44full}, one
sees that electrons must maintain an index $q_{e}\approx1.75$ between
$0.46<E<5.8$ GeV. While the upper bound is acceptable given the spectral
break proton energy inferred from the super GeV emission measured
by the Fermi LAT, the lower end is rather uncomfortable, since the
nonlinear modification of both protons and electrons with the Bohm
(or other similar for protons and electrons turbulent diffusivities)
starts (slowly) only at the proton rest energy. The calculated nonlinear
spectra are shown in Fig.\ref{fig:Spectra-of-protons} for the both
species. At and below $1$GeV, the electron spectrum is very close
to the test particle solution, $q_{e}\approx2,$ even though the proton
spectrum may be somewhat steeper there, as we mentioned above.

\paragraph*{Photon spectra}

\begin{figure}
\includegraphics[scale=0.7]{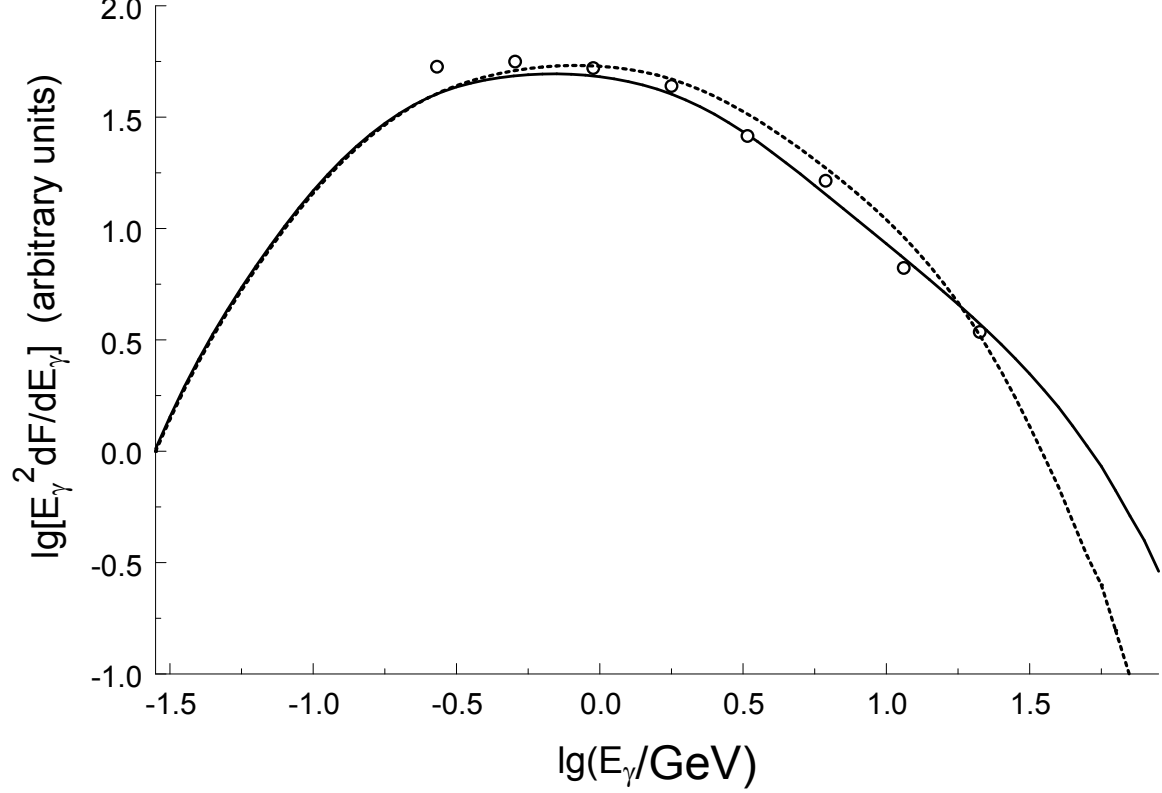}

\caption{\textbf{Gamma radiation spectra.} Photon spectra resulting from $\pi^{0}$
decay and calculated for two different parent proton spectra compared
against the Fermi data (open circles). Solid line: a test particle
acceleration regime with the spectral index $q=2$ below the break
and $q=3$ above the break at $p_{br}=7GeV$/c. Dashed line: a moderately
nonlinear acceleration regime corresponding to the spectrum shown
in Fig.\ref{fig:Spectra-of-protons} ($q\simeq1.75$ and $q\simeq2.75$
below and above the break, respectively). Cut-offs are placed at 300
GeV for TP- and 100 GeV, for NL-spectrum. \label{fig:Gamma-emission-from}}

\end{figure}

The above considerations somewhat weaken the radio data as a probe
for the slope of the electron and (more importantly) for the proton
spectrum. Therefore, the exact degree of nonlinearity of the acceleration
remains unknown and we consider both the TP and weakly NL regimes
in our calculations of the photon spectra, generated in $p-p$ collisions.
The results are shown in Fig.\ref{fig:Gamma-emission-from}. The best
fit to the Fermi data is provided by a TP energy distribution ($\propto E^{-2}$)
below $p_{br}\simeq7$GeV/c with the spectrum steepening by exactly
one power above it. Note that the small deviation of the computed
spectrum from the lowest energy \emph{Fermi-}point may also be corrected
by including the Bremsstrahlung of the secondary electrons \citep{Uchiyama10}.
The spectrum steepening is perfectly consistent with the proton partial
escape described above (with no parameters involved) and shown in
Fig.\ref{fig:Scattering-zone-of}. For comparison, a weakly NL spectrum,
shown in Fig.\ref{fig:Spectra-of-protons}, is also used for these
calculations (dashed line in Fig.\ref{fig:Gamma-emission-from}),
but its fit would require a stronger break ($\Delta q>1$) or a low
momentum cut-off, Fig.\ref{fig:Gamma-emission-from}, i.e. at least
one additional free parameter.

\section*{Discussion}

To summarize the results, the previously suggested mechanism for a
break on the spectrum of shock accelerated protons \citep{MDS05}
is found consistent with the recent \citep{Abdo10W44full} Fermi LAT
observations of the SNR W44. The observed gamma ray spectrum most
likely results from the decay of $\pi^{0}$-mesons which are born
in $p-p$ collisions of shock accelerated protons with an ambient
dense gas. The parent proton spectrum is best represented by a classical
test particle power law $\propto E^{-2}$, steepening to $E^{-3}$
at $E_{br}\approx7GeV$ due to deteriorated particle confinement caused
by the ion-neutral collisions and the resultant Alfven wave evanescence.
The position of the break momentum in the particle spectrum may be
estimated using eq.(\ref{eq:p1}), or conversely, the combination
of parameters involved in this estimate can be inferred from the measured
break momentum. The cut-off momentum is not constrained in this scenario. 

An alternative explanation, based on a different mechanism of the
break, associated with the change of the particle transport in the
CR shock precursor \citep{MD06} is also possible but is less definitive
in the spectrum slope variation $\Delta q$ across the break (see
also \citep{Uchiyama10} for the most recent alternative suggestions).
In addition, the ref \citep{MD06} mechanism would imply a considerable
nonlinearity, i.e. a stronger CR shock precursor than it seems appropriate
for the observed low break momentum. Still alternatively, assuming
our {}``environmental'' break mechanism is at work, i.e. $\Delta q=1$,
but the shock structure appreciably modified, we arrive at the $E^{-1.75}$
spectrum below the break (as the radio observations suggest for the
electrons), and $E^{-2.75}$ above the break. A fit to the data is
marginally possible, but it would require a relatively low cut-off
momentum at about $100$ GeV/c. This possibility may be supported
or ruled out once the data (upper limit) around this energy become
available. 

The most robust and attractive aspect of the suggested mechanism for
the spectral break is the exact $\Delta q=1$ variation of the spectral
index. Indeed, this change in the spectral slope is due to truncation
of the particle momentum phase space and does not depend on any parameters.
In a combination with the test particle regime operating below the
break, which is consistent with the low value of the break momentum,
it provides a very good fit to the \emph{Fermi }LAT data. Of three
physically different types of spectral breaks suggested earlier \citep{mdj02,MD06,MDS05},
namely the current, {}``environmental'' mechanism appears to be
plausible in situations where dense target gas is present, as required
for efficient $\pi^{0}$ production. In a more general context, a
spectral break is a natural resolution of the well known but puzzling
trend of the \emph{nonlinear }(i.e. supposedly improved) DSA theory
to develop spectra which are even harder than the simple test particle
spectra, thus encountering more difficulties while accommodating the
bulk of observations \citep{Gaisser98,Hillas05}. Such a spectrum
-- i.e., diverging in energy-- exhausts the shock energy available
for the acceleration as the cut-off momentum grows. Note, that the
spectrum of the RX J1713.7-3946 \citep{Ahar06RXJ} is also consistent
with the same break mechanism but with $p_{br}\sim10^{3}GeV/c$ and
with naturally stronger acceleration nonlinearity \citep{MDS05}.
However, it is difficult to make the case for hadronic origin of the
gamma-ray emission of the RX J1713.7-3946 \citep{AharNat04,Ahar06RXJ,WaxmanRXJ08}.
The fundamental role of the W44 for the problem of CR origin is that
this particular remnant seems to rule out contaminating electron emission
due to Bremsstrahlung and the inverse Compton scattering \citep{Abdo10W44full,Uchiyama10}
thus favoring the hadronic origin of the gamma emission.

\section*{Methods}

\paragraph*{Acceleration model and proton spectrum}

Methods of calculation of the accelerated particle spectra, including
the particle back-reaction on the shock structure, are now well developed.
We use the diffusion-convection equation for describing the distribution
of high energy particles (CRs). To include the back-reaction, three
further relations are used in a quasi-stationary acceleration regime.
These are the conservation of mass and momentum fluxes in the smooth
part of the shock transition (so called CR-precursor) and the Rankine-Hugoniot
relation for the shock compression at the gaseous discontinuity (sub-shock).
The complete system of equations is then reduced to one nonlinear
integral equation \citep{MDru01}. This equation is solved numerically
and the results are shown in Fig.\ref{fig:Spectra-of-protons}. The
above method of calculation of the parent particle spectra is shown
to be in excellent agreement \citep{Mosk08} with numerical simulations
\citep{BerezhEllis99,EllisBerBar00} as well as with other semi-analytic
approaches \citep{Blasi2005}.

\paragraph*{Gamma-ray emission spectrum}

Once the parent proton spectrum is obtained, we calculate the $\pi^{0}$
production rate and the gamma-ray emissivity. In so doing, we adopt
numerical recipe described in detail in \citep{Kamae06,KarlssonKamae08}.
The physical processes behind these calculations are \citep{DermerPPcol86}
(i) collisions of accelerated protons with the protons of the ambient
gas resulting in the following spectrum of $\pi^{0}$-mesons:

\begin{equation}
F_{pp}\left(E_{\pi}\right)=4\pi N_{pg}\intop\frac{d\sigma\left(E_{\pi},E_{p}\right)}{dE_{\pi}}J_{p}\left(E_{p}\right)dE_{p}\label{eq:pionsp}\end{equation}
where, $N_{pg}$ is the number density of protons in the gas, $d\sigma/dE_{\pi}$
is the differential cross section for the $\pi^{0}$ production in
collisions between accelerated protons of energy $E_{p}$ and gas
protons, $J_{p}$ is the flux of accelerated protons, $E_{\pi}$ is
the energy of $\pi^{0}$ mesons; (ii) decay of $\pi^{0}$ resulting
in the gamma emission spectrum

\begin{equation}
F\left(E_{\gamma}\right)=2\intop_{E_{\gamma}+m_{\pi}^{2}c^{4}/4E_{\gamma}}^{\infty}\frac{F_{pp}\left(E_{\pi}\right)}{\sqrt{E_{\pi}^{2}-m_{\pi}^{2}c^{4}}}dE_{\pi}\label{eq:gammasp}\end{equation}
where $m_{\pi}$ is the pion rest mass.

\subsection*{Acknowledgments}

We wold like to thank Sabrina Casanova for furnishing a detailed information
about the code developed in ref.\citep{KarlssonKamae08}. Support
by NASA under the Grants NNX 07AG83G and NNX09AT94G as well as by
the Department of Energy, Grant No. DE-FG02-04ER54738 is gratefully
acknowledged.

\subsection*{Author contributions}

M.A.M. carried out the calculations of the spectra, P.H.D. contributed
to the wave propagation part, R.Z.S. contributed to the wave-particle
interaction part. The article was written by M.A.M and edited by all
the co-authors.

\subsection*{Corresponding author}

M.A. Malkov, e-mail: mmalkov@ucsd.edu

\paragraph*{Competing financial interests}

The authors declare no competing financial interests.

\pagebreak

\begin{thebibliography}{10}
\expandafter\ifx\csname url\endcsname\relax
  \def\url#1{\texttt{#1}}\fi
\expandafter\ifx\csname urlprefix\endcsname\relax\def\urlprefix{URL }\fi
\providecommand{\bibinfo}[2]{#2}
\providecommand{\eprint}[2][]{\url{#2}}

\bibitem{Reynolds08ARA}
\bibinfo{author}{{Reynolds}, S.~P.}
\newblock \bibinfo{title}{{Supernova Remnants at High Energy}}.
\newblock \emph{\bibinfo{journal}{\araa}} \textbf{\bibinfo{volume}{46}},
  \bibinfo{pages}{89--126} (\bibinfo{year}{2008}).

\bibitem{Raymond01}
\bibinfo{author}{{Raymond}, J.~C.}
\newblock \bibinfo{title}{{Optical and UV Diagnostics of Supernova Remnant
  Shocks}}.
\newblock \emph{\bibinfo{journal}{Space Science Reviews}}
  \textbf{\bibinfo{volume}{99}}, \bibinfo{pages}{209--218}
  (\bibinfo{year}{2001}).

\bibitem{Gaisser98}
\bibinfo{author}{{Gaisser}, T.~K.}, \bibinfo{author}{{Protheroe}, R.~J.} \&
  \bibinfo{author}{{Stanev}, T.}
\newblock \bibinfo{title}{{Gamma-Ray Production in Supernova Remnants}}.
\newblock \emph{\bibinfo{journal}{\apj}} \textbf{\bibinfo{volume}{492}},
  \bibinfo{pages}{219--227} (\bibinfo{year}{1998}).

\bibitem{Hillas05}
\bibinfo{author}{{Hillas}, A.~M.}
\newblock \bibinfo{title}{{TOPICAL REVIEW: Can diffusive shock acceleration in
  supernova remnants account for high-energy galactic cosmic rays?}}
\newblock \emph{\bibinfo{journal}{Journal of Physics G Nuclear Physics}}
  \textbf{\bibinfo{volume}{31}}, \bibinfo{pages}{95--131}
  (\bibinfo{year}{2005}).

\bibitem{Ahar06RXJ}
\bibinfo{author}{{Aharonian}, F.} \emph{et~al.}
\newblock \bibinfo{title}{{A detailed spectral and morphological study of the
  gamma-ray supernova remnant RX J1713.7-3946 with HESS}}.
\newblock \emph{\bibinfo{journal}{\aap}} \textbf{\bibinfo{volume}{449}},
  \bibinfo{pages}{223--242} (\bibinfo{year}{2006}).

\bibitem{Abdo10W44full}
\bibinfo{author}{{Abdo}, A.~A.} \emph{et~al.}
\newblock \bibinfo{title}{{Gamma-Ray Emission from the Shell of Supernova
  Remnant W44 Revealed by the Fermi LAT}}.
\newblock \emph{\bibinfo{journal}{Science}} \textbf{\bibinfo{volume}{327}},
  \bibinfo{pages}{1103--1106} (\bibinfo{year}{2010}).

\bibitem{Enomoto02}
\bibinfo{author}{{Enomoto}, R.} \emph{et~al.}
\newblock \bibinfo{title}{{The acceleration of cosmic-ray protons in the
  supernova remnant RX J1713.7-3946}}.
\newblock \emph{\bibinfo{journal}{\nat}} \textbf{\bibinfo{volume}{416}},
  \bibinfo{pages}{823--826} (\bibinfo{year}{2002}).

\bibitem{Pohl02}
\bibinfo{author}{{Reimer}, O.} \& \bibinfo{author}{{Pohl}, M.}
\newblock \bibinfo{title}{{No evidence yet for hadronic TeV gamma-ray emission
  from SNR RX J1713.7-3946}}.
\newblock \emph{\bibinfo{journal}{\aap}} \textbf{\bibinfo{volume}{390}},
  \bibinfo{pages}{L43--L46} (\bibinfo{year}{2002}).

\bibitem{MDS05}
\bibinfo{author}{{Malkov}, M.~A.}, \bibinfo{author}{{Diamond}, P.~H.} \&
  \bibinfo{author}{{Sagdeev}, R.~Z.}
\newblock \bibinfo{title}{{On the Gamma-Ray Spectra Radiated by Protons
  Accelerated in Supernova Remnant Shocks near Molecular Clouds: The case of
  Supernova Remnant RX J1713.7-3946}}.
\newblock \emph{\bibinfo{journal}{\apjl}} \textbf{\bibinfo{volume}{624}},
  \bibinfo{pages}{L37--L40} (\bibinfo{year}{2005}).

\bibitem{Drury83}
\bibinfo{author}{{Drury}, L.~O.}
\newblock \bibinfo{title}{{An introduction to the theory of diffusive shock
  acceleration of energetic particles in tenuous plasmas}}.
\newblock \emph{\bibinfo{journal}{Reports on Progress in Physics}}
  \textbf{\bibinfo{volume}{46}}, \bibinfo{pages}{973--1027}
  (\bibinfo{year}{1983}).

\bibitem{BlandEich87}
\bibinfo{author}{{Blandford}, R.} \& \bibinfo{author}{{Eichler}, D.}
\newblock \bibinfo{title}{{Particle Acceleration at Astrophysical Shocks - a
  Theory of Cosmic-Ray Origin}}.
\newblock \emph{\bibinfo{journal}{\physrep}} \textbf{\bibinfo{volume}{154}},
  \bibinfo{pages}{1--75} (\bibinfo{year}{1987}).

\bibitem{KulsrNeutr69}
\bibinfo{author}{{Kulsrud}, R.} \& \bibinfo{author}{{Pearce}, W.~P.}
\newblock \bibinfo{title}{{The Effect of Wave-Particle Interactions on the
  Propagation of Cosmic Rays}}.
\newblock \emph{\bibinfo{journal}{\apj}} \textbf{\bibinfo{volume}{156}},
  \bibinfo{pages}{445--469} (\bibinfo{year}{1969}).

\bibitem{ZweibelShull82}
\bibinfo{author}{{Zweibel}, E.~G.} \& \bibinfo{author}{{Shull}, J.~M.}
\newblock \bibinfo{title}{{Confinement of cosmic rays in molecular clouds}}.
\newblock \emph{\bibinfo{journal}{\apj}} \textbf{\bibinfo{volume}{259}},
  \bibinfo{pages}{859--868} (\bibinfo{year}{1982}).

\bibitem{VoelkNeutrDamp81}
\bibinfo{author}{{Voelk}, H.~J.}, \bibinfo{author}{{Morfill}, G.~E.} \&
  \bibinfo{author}{{Forman}, M.~A.}
\newblock \bibinfo{title}{{The effect of losses on acceleration of energetic
  particles by diffusive scattering through shock waves}}.
\newblock \emph{\bibinfo{journal}{\apj}} \textbf{\bibinfo{volume}{249}},
  \bibinfo{pages}{161--175} (\bibinfo{year}{1981}).

\bibitem{DruryNeutral96}
\bibinfo{author}{{Drury}, L.~O.~C.}, \bibinfo{author}{{Duffy}, P.} \&
  \bibinfo{author}{{Kirk}, J.~G.}
\newblock \bibinfo{title}{{Limits on diffusive shock acceleration in dense and
  incompletely ionised media.}}
\newblock \emph{\bibinfo{journal}{\aap}} \textbf{\bibinfo{volume}{309}},
  \bibinfo{pages}{1002--1010} (\bibinfo{year}{1996}).

\bibitem{RevilleNeutr08}
\bibinfo{author}{{Reville}, B.}, \bibinfo{author}{{Kirk}, J.~G.},
  \bibinfo{author}{{Duffy}, P.} \& \bibinfo{author}{{O'Sullivan}, S.}
\newblock \bibinfo{title}{{Environmental Limits on the Nonresonant Cosmic-Ray
  Current-Driven Instability}}.
\newblock \emph{\bibinfo{journal}{International Journal of Modern Physics D}}
  \textbf{\bibinfo{volume}{17}}, \bibinfo{pages}{1795--1801}
  (\bibinfo{year}{2008}).

\bibitem{CrutcherMC99}
\bibinfo{author}{{Crutcher}, R.~M.}
\newblock \bibinfo{title}{{Magnetic Fields in Molecular Clouds: Observations
  Confront Theory}}.
\newblock \emph{\bibinfo{journal}{\apj}} \textbf{\bibinfo{volume}{520}},
  \bibinfo{pages}{706--713} (\bibinfo{year}{1999}).

\bibitem{Chev03}
\bibinfo{author}{{Chevalier}, R.~A.} \& \bibinfo{author}{{Oishi}, J.}
\newblock \bibinfo{title}{{Cassiopeia A and Its Clumpy Presupernova Wind}}.
\newblock \emph{\bibinfo{journal}{\apjl}} \textbf{\bibinfo{volume}{593}},
  \bibinfo{pages}{L23--L26} (\bibinfo{year}{2003}).

\bibitem{Pariz04}
\bibinfo{author}{{Parizot}, E.}, \bibinfo{author}{{Marcowith}, A.},
  \bibinfo{author}{{van der Swaluw}, E.}, \bibinfo{author}{{Bykov}, A.~M.} \&
  \bibinfo{author}{{Tatischeff}, V.}
\newblock \bibinfo{title}{{Superbubbles and energetic particles in the Galaxy.
  I. Collective effects of particle acceleration}}.
\newblock \emph{\bibinfo{journal}{\aap}} \textbf{\bibinfo{volume}{424}},
  \bibinfo{pages}{747--760} (\bibinfo{year}{2004}).

\bibitem{DraineMcKee93}
\bibinfo{author}{{Draine}, B.~T.} \& \bibinfo{author}{{McKee}, C.~F.}
\newblock \bibinfo{title}{{Theory of interstellar shocks}}.
\newblock \emph{\bibinfo{journal}{\araa}} \textbf{\bibinfo{volume}{31}},
  \bibinfo{pages}{373--432} (\bibinfo{year}{1993}).

\bibitem{NakanoMC84}
\bibinfo{author}{{Nakano}, T.}
\newblock \bibinfo{title}{{Contraction of magnetic interstellar clouds}}.
\newblock \emph{\bibinfo{journal}{Fundamentals of Cosmic Physics}}
  \textbf{\bibinfo{volume}{9}}, \bibinfo{pages}{139--231}
  (\bibinfo{year}{1984}).

\bibitem{ShullMcKeeMC79}
\bibinfo{author}{{Shull}, J.~M.} \& \bibinfo{author}{{McKee}, C.~F.}
\newblock \bibinfo{title}{{Theoretical models of interstellar shocks. I -
  Radiative transfer and UV precursors}}.
\newblock \emph{\bibinfo{journal}{\apj}} \textbf{\bibinfo{volume}{227}},
  \bibinfo{pages}{131--149} (\bibinfo{year}{1979}).

\bibitem{BykovMC00}
\bibinfo{author}{{Bykov}, A.~M.}, \bibinfo{author}{{Chevalier}, R.~A.},
  \bibinfo{author}{{Ellison}, D.~C.} \& \bibinfo{author}{{Uvarov}, Y.~A.}
\newblock \bibinfo{title}{{Nonthermal Emission from a Supernova Remnant in a
  Molecular Cloud}}.
\newblock \emph{\bibinfo{journal}{\apj}} \textbf{\bibinfo{volume}{538}},
  \bibinfo{pages}{203--216} (\bibinfo{year}{2000}).

\bibitem{Uchiyama10}
\bibinfo{author}{{Uchiyama}, Y.}, \bibinfo{author}{{Blandford}, R.~D.},
  \bibinfo{author}{{Funk}, S.}, \bibinfo{author}{{Tajima}, H.} \&
  \bibinfo{author}{{Tanaka}, T.}
\newblock \bibinfo{title}{{Gamma-ray Emission from Crushed Clouds in Supernova
  Remnants}}.
\newblock \emph{\bibinfo{journal}{\apjl}} \textbf{\bibinfo{volume}{723}},
  \bibinfo{pages}{L122--L126} (\bibinfo{year}{2010}).

\bibitem{Reach05}
\bibinfo{author}{{Reach}, W.~T.}, \bibinfo{author}{{Rho}, J.} \&
  \bibinfo{author}{{Jarrett}, T.~H.}
\newblock \bibinfo{title}{{Shocked Molecular Gas in the Supernova Remnants W28
  and W44: Near-Infrared and Millimeter-Wave Observations}}.
\newblock \emph{\bibinfo{journal}{\apj}} \textbf{\bibinfo{volume}{618}},
  \bibinfo{pages}{297--320} (\bibinfo{year}{2005}).

\bibitem{W44Radio07}
\bibinfo{author}{{Castelletti}, G.}, \bibinfo{author}{{Dubner}, G.},
  \bibinfo{author}{{Brogan}, C.} \& \bibinfo{author}{{Kassim}, N.~E.}
\newblock \bibinfo{title}{{The low-frequency radio emission and spectrum of the
  extended SNR W44: new VLA observations at 74 and 324 MHz}}.
\newblock \emph{\bibinfo{journal}{\aap}} \textbf{\bibinfo{volume}{471}},
  \bibinfo{pages}{537--549} (\bibinfo{year}{2007}).

\bibitem{MD06}
\bibinfo{author}{{Malkov}, M.~A.} \& \bibinfo{author}{{Diamond}, P.~H.}
\newblock \bibinfo{title}{{Nonlinear Shock Acceleration Beyond the Bohm
  Limit}}.
\newblock \emph{\bibinfo{journal}{\apj}} \textbf{\bibinfo{volume}{642}},
  \bibinfo{pages}{244--259} (\bibinfo{year}{2006}).

\bibitem{mdj02}
\bibinfo{author}{{Malkov}, M.~A.}, \bibinfo{author}{{Diamond}, P.~H.} \&
  \bibinfo{author}{{Jones}, T.~W.}
\newblock \bibinfo{title}{{On the Possible Reason for Nondetection of TeV
  Protons in Supernova Remnants}}.
\newblock \emph{\bibinfo{journal}{\apj}} \textbf{\bibinfo{volume}{571}},
  \bibinfo{pages}{856--865} (\bibinfo{year}{2002}).

\bibitem{AharNat04}
\bibinfo{author}{{Aharonian}, F.~A.} \emph{et~al.}
\newblock \bibinfo{title}{{High-energy particle acceleration in the shell of a
  supernova remnant}}.
\newblock \emph{\bibinfo{journal}{\nat}} \textbf{\bibinfo{volume}{432}},
  \bibinfo{pages}{75--77} (\bibinfo{year}{2004}).

\bibitem{WaxmanRXJ08}
\bibinfo{author}{{Katz}, B.} \& \bibinfo{author}{{Waxman}, E.}
\newblock \bibinfo{title}{{In which shell-type SNRs should we look for
  gamma-rays and neutrinos from P P collisions?}}
\newblock \emph{\bibinfo{journal}{Journal of Cosmology and Astro-Particle
  Physics}} \textbf{\bibinfo{volume}{1}}, \bibinfo{pages}{1--29}
  (\bibinfo{year}{2008}).

\bibitem{MDru01}
\bibinfo{author}{{Malkov}, M.~A.} \& \bibinfo{author}{{Drury}, L.~O.}
\newblock \bibinfo{title}{{Nonlinear theory of diffusive acceleration of
  particles by shock waves }}.
\newblock \emph{\bibinfo{journal}{Reports on Progress in Physics}}
  \textbf{\bibinfo{volume}{64}}, \bibinfo{pages}{429--481}
  (\bibinfo{year}{2001}).

\bibitem{Mosk08}
\bibinfo{author}{{Moskalenko}, I.~V.}, \bibinfo{author}{{Porter}, T.~A.},
  \bibinfo{author}{{Malkov}, M.~A.} \& \bibinfo{author}{{Diamond}, P.~H.}
\newblock \bibinfo{title}{{Hadronic Gamma Rays from Supernova Remnants}}.
\newblock In \emph{\bibinfo{booktitle}{International Cosmic Ray Conference}},
  vol.~\bibinfo{volume}{2} of \emph{\bibinfo{series}{International Cosmic Ray
  Conference}}, \bibinfo{pages}{763--766} (\bibinfo{year}{2008}).

\bibitem{BerezhEllis99}
\bibinfo{author}{{Berezhko}, E.~G.} \& \bibinfo{author}{{Ellison}, D.~C.}
\newblock \bibinfo{title}{{A Simple Model of Nonlinear Diffusive Shock
  Acceleration}}.
\newblock \emph{\bibinfo{journal}{\apj}} \textbf{\bibinfo{volume}{526}},
  \bibinfo{pages}{385--399} (\bibinfo{year}{1999}).

\bibitem{EllisBerBar00}
\bibinfo{author}{{Ellison}, D.~C.}, \bibinfo{author}{{Berezhko}, E.~G.} \&
  \bibinfo{author}{{Baring}, M.~G.}
\newblock \bibinfo{title}{{Nonlinear Shock Acceleration and Photon Emission in
  Supernova Remnants}}.
\newblock \emph{\bibinfo{journal}{\apj}} \textbf{\bibinfo{volume}{540}},
  \bibinfo{pages}{292--307} (\bibinfo{year}{2000}).

\bibitem{Blasi2005}
\bibinfo{author}{{Blasi}, P.}, \bibinfo{author}{{Gabici}, S.} \&
  \bibinfo{author}{{Vannoni}, G.}
\newblock \bibinfo{title}{{On the role of injection in kinetic approaches to
  non-linear particle acceleration at non-relativistic shock waves}}.
\newblock \emph{\bibinfo{journal}{\mnras}} \textbf{\bibinfo{volume}{361}},
  \bibinfo{pages}{907--918} (\bibinfo{year}{2005}).

\bibitem{Kamae06}
\bibinfo{author}{{Kamae}, T.}, \bibinfo{author}{{Karlsson}, N.},
  \bibinfo{author}{{Mizuno}, T.}, \bibinfo{author}{{Abe}, T.} \&
  \bibinfo{author}{{Koi}, T.}
\newblock \bibinfo{title}{{Parameterization of {gamma}, e{+/-}, and Neutrino
  Spectra Produced by p-p Interaction in Astronomical Environments}}.
\newblock \emph{\bibinfo{journal}{\apj}} \textbf{\bibinfo{volume}{647}},
  \bibinfo{pages}{692--708} (\bibinfo{year}{2006}).

\bibitem{KarlssonKamae08}
\bibinfo{author}{{Karlsson}, N.} \& \bibinfo{author}{{Kamae}, T.}
\newblock \bibinfo{title}{{Parameterization of the Angular Distribution of
  Gamma Rays Produced by p-p Interaction in Astronomical Environments}}.
\newblock \emph{\bibinfo{journal}{\apj}} \textbf{\bibinfo{volume}{674}},
  \bibinfo{pages}{278--285} (\bibinfo{year}{2008}).

\bibitem{DermerPPcol86}
\bibinfo{author}{{Dermer}, C.~D.}
\newblock \bibinfo{title}{{Secondary production of neutral pi-mesons and the
  diffuse galactic gamma radiation}}.
\newblock \emph{\bibinfo{journal}{\aap}} \textbf{\bibinfo{volume}{157}},
  \bibinfo{pages}{223--229} (\bibinfo{year}{1986}).

\end{thebibliography}
\end{document}